\documentclass[amsmath,amssymb,aps,twocolumn,prb,longbibliography,floatfix,superscriptaddress]{revtex4-2}
\usepackage{graphicx,bm,times}
\usepackage[utf8]{inputenc}
\usepackage{braket}
\usepackage{amsfonts}
\usepackage{color}
\usepackage{chngcntr}

\usepackage[breaklinks]{hyperref}
\hypersetup{
    pdfstartview={FitH},
    colorlinks=true,
    linkcolor=blue,
    citecolor=blue,
    urlcolor=blue
}

\begin{document}

\title{Composition-driven Mott transition within SrTi$_{\rm 1-x}$V$_{\rm x}$O$_3$}

\author{A.~D.~N.~James}
\affiliation{H.~H.~Wills Physics Laboratory,
University of Bristol, Tyndall Avenue, Bristol, BS8 1TL, United Kingdom}

\author{M.~Aichhorn}
\affiliation{Institute of Theoretical and Computational Physics, TU Graz, NAWI
Graz, Petersgasse 16, 8010 Graz, Austria}

\author{J.~Laverock}
\affiliation{School of Chemistry,
University of Bristol, Cantocks Close, Bristol, BS8 1TS, United Kingdom}

\date{\today}

\begin{abstract}
The last few decades has seen the rapid growth of interest in the bulk perovskite-type transition metal oxides SrVO$_3$ and SrTiO$_3$. The electronic configuration of these perovskites differs by one electron associated to the transition metal species which gives rise to the drastically different electronic properties. Therefore, it is natural to look into how the electronic structure transitions between these bulk structures by using doping. Measurements of the substitutional doped SrTi$_{\rm 1-x}$V$_{\rm x}$O$_3$ shows an metal-insulator transition (MIT) as a function of doping. By using supercell density functional theory with dynamical mean field theory (DFT+DMFT), we show that the MIT is indeed the result of the combination of local electron correlation effects (Mott physics) within the t$_{\rm 2g}$ orbitals and the atomic site configuration of the transition metals which may indicate dependence on site disorder. SrTi$_{\rm 1-x}$V$_{\rm x}$O$_3$ may be an ideal candidate for benchmarking cutting-edge Mott-Anderson models of real systems. We show that applying an effective external perturbation on SrTi$_{\rm 1-x}$V$_{\rm x}$O$_3$ can switch the system between the insulating and metallic phase, meaning this is a bulk system with the potential use in Mott electronic devices.

\end{abstract}

\maketitle

\section{Introduction}
There is continuing interest in oxide based materials as they exhibit exotic phenomena, such as colossal
magnetoresistance; high-temperature superconductivity; and Mott insulating
phases, or respond to external stimuli which is highly desirable for innovating electronic devices~\cite{Ha_2011,mannhart2010,brahlek2017}. 
In particular, perovskite-type transition metal oxides ABO$_3$ show many desirable properties such as a metal-insulator transition (MIT)~\cite{Imada_1998} which are of interest for developing Mott devices~\cite{newns1998, brahlek2017}. It is the many-body correlated behaviour of electrons in these materials which gives rise to the emergence of these exotic phenomena. Here, the interaction energy between electrons is comparable to their kinetic energy. Therefore, it is important to understand the correlated behaviour of electrons and how to manipulate them for future technological applications. 

Bulk cubic SrVO$_3$ is a prototypical correlated material \cite{nekrasov2005} 
and is thought to have several applications owing to its high performance ~\cite{Zhang-2016,Mirjolet_2019,brahlek2017}. It has a three peak spectral structure around the Fermi level with the central part consisting of sharp $3d$ quasiparticle bands at low excitation energies giving a well-defined Fermi surface \cite{yoshida2010,aizaki2012}. These quasiparticle bands are accompanied by localised states (the so called incoherent Hubbard sidebands) at an energy scale comparable with the Coulomb repulsion parameter, $U$~\cite{sekiyama2004,pen1999,laverock2013b}. This material has been the subject of many density functional theory with dynamical mean field theory (DFT+DMFT) studies \cite{nekrasov2005,sekiyama2004,nekrasov2006,kotliar2006,byczuk2007,tomczak2012}, 
where only one electron exists within the V $3d$ $t_{\rm 2g}$. Often only the V $3d$ $t_{\rm 2g}$ states around the Fermi level are used in the DMFT calculations as the $e_{\rm g}$ do not hybridise with the $t_{\rm 2g}$ states and are also unoccupied. DFT+DMFT is able to describe all of the on-site local correlations \cite{georges1996,kotliar2006}, and has been well-tested on SrVO$_3$ with very good results, including the energetics and spectral weight of Hubbard sidebands and QP renormalisation~\cite{sekiyama2004,nekrasov2006,kotliar2006,byczuk2007,tomczak2012}. However, $GW$+DMFT~\cite{Boehnke2016} predictions reinterprets SrVO$_3$ as a weakly correlated material with low static local interactions, since their results show pronounced plasmonic satellites due to screening.

The importance of electron correlations in a system may be gauged in terms of the ratio, $U/W$, where $W$ is the non-interacting bandwidth as predicted by DFT. There are certain factors which been shown to have influenced this ratio. Crystal field effects have been a prime example of this as it can reduce $W$ to give the Mott insulating phase as predicted by DFT+DMFT of bilayer capped SrVO$_3$~\cite{zhong2015} which agrees with that seen in experimental spectral data of SrVO$_3$ thin films (below approximately 6 unit cells)~\cite{yoshimatsu2010,gu2014}. These effects are also predicted in other vanadates where these can be tuned by strain~\cite{bhandary2016,schuler2018,hampel2020,beck2018,sclauzero2016}. Conversely, $W$ can also be influenced by quantum confinement of heterostructure superlattices~\cite{james2021quantum} in agreement with the corresponding experimental data~\cite{laverock2017,gu2018,kobayashi2015,kobayashi2017}.

Bulk SrTiO$_3$ is a band-insulator despite sharing  the same cubic structure as SrVO$_3$ at room temperature (below $\sim$ 105 K, SrTiO$_3$ exhibits a tetragonal structural transition \cite{ROSSELLA200795}). SrTiO$_3$ has been the focus of modern day research due to its many desirable properties which can be exploited for developing electronic devices ~\cite{Ha_2011,Li_2019,xu2022review,Kleemann_2020,SHI2020105195}. 
The main interest in SrTiO$_3$ is it often being used as a key building block for new oxide heterostructures either as a substrate to grow other oxide compounds on or within oxide heterostuctures which leads to unique phenomena originating from the interface between oxides (examples of which are discussed in Refs.~\cite{Yan_2018,Pai_2018,Christensen_2019}). 
Evidently, it is this versatility of SrTiO$_3$ which has resulted in its own field of research. 

It is interesting how these two bulk (structurally simple) perovskite transition metal oxides compounds where the transition metal species effectively differs by an electron (and a proton) give rise to such different electronic structures. 
Therefore, tuning the electronic structure between SrVO$_3$ and SrTiO$_3$,  achieved by introducing impurities or defects in the material, will give an insight into the rise of interesting correlated electron behaviour. For example, 
substitutional doping has been shown to be an effective way to introduce filling-controlled MIT in the aliovalent A-site substitution of the intensely studied La$_{\rm 1-x}$Sr$_{\rm x}$VO$_3$~\cite{Miyasaka_2000,Takahashi_2022} 
(where $\rm x$ represents doping concentration). 
Recent studies of B-site substitutionally doped SrTi$_{\rm 1-x}$V$_{\rm x}$O$_3$ show that there is a temperature-dependent MIT which is composition driven~\cite{HONG2002305,Kanda_2021,Gu_2013_STVO,Tsuiki_1983}. Beyond this composition-driven MIT, other properties of this doped system have been investigated such as its impedance spectroscopy and electrical
conduction mechanism for certain dopings~\cite{mantry2019investigation} and it may also be classified as a correlated transparent conductive oxide~\cite{Kanda_2021} which has the potential for use in optoelectronic devices. 
Several studies have measured the critical doping for the MIT to be at different dopings, ranging from between 0.4~\cite{Kanda_2021}, 0.6~\cite{Tsuiki_1983}, 0.67~\cite{Gu_2013_STVO} and 0.7~\cite{HONG2002305}. The critical doping was shown to be restrained between 0.5 and 1 when introducing oxygen vacancies~\cite{ITAKA2005222}. 

The driving force behind the MIT in SrTi$_{\rm 1-x}$V$_{\rm x}$O$_3$ is still not fully understood with speculation of its mechanism. Tsuiki $\textit{et al.}$ interpreted their findings by arguing that the distortion from the Jahn-Teller effect~\cite{STURGE196891} makes the donor levels so deep that the vanadium $d$-orbitals cannot overlap to form a band~\cite{Tsuiki_1983}. Hong $\textit{et al.}$ proposed that this is MIT is controlled by the Anderson localized states in which a MIT occurs where the mobility edge crosses the Fermi level~\cite{HONG2002305}. Subsequent studies then proposed that the mechanism behind this composition drived MIT is due to a combination of both electron correlation effects (Mott physics) and disorder-induced (Anderson) localisation~\cite{Kanda_2021,Gu_2013_STVO}, where the Ti$^{4+}$ ion substitution introduces Anderson-localized states along with lattice distortions which result in a reduction in the effective bandwidth $W$. Therefore, it is clearly necessary to understand how the electronic structure transitions in SrTi$_{\rm 1-x}$V$_{\rm x}$O$_3$ between the band-insulator to the correlated metal from a theoretical perspective in order for this doped system to be effectively manipulated. 

Here, we look at the origin of the dominant mechanism which gives rise to the MIT in SrTi$_{\rm 1-x}$V$_{\rm x}$O$_3$. We focus on using the DFT+DMFT method on supercells of different configurations of the highly symmetric atomic positions in order to gain insight into the mechanism. We find that from the set of configurations studied here, the MIT is highly dependent on both the local environment of the V atoms in the doped systems (which is influenced by the site configuration) and the Mott local electron correlation effects within the t$_{\rm 2g}$ orbitals. 
The doping-dependent non-interacting $W$ in combination with the on-site $U$ pushes the system, depending on the $U$/$W$ ratio, into either the metallic (low $U$/$W$) or insulating phase (high $U$/$W$). 
Around the MIT critical doping, the MIT appears to be sensitive to crystal field effects on $W$ introduced from site configuration of the transition metal species within the unit cells. Finally, we show that the effect of an external perturbation on Mott insulating dopings can push them metallic which has potential applications for Mott devices. 

\section{Method}

\begin{figure}[t!]
 \centerline{\includegraphics[width=0.995\linewidth]{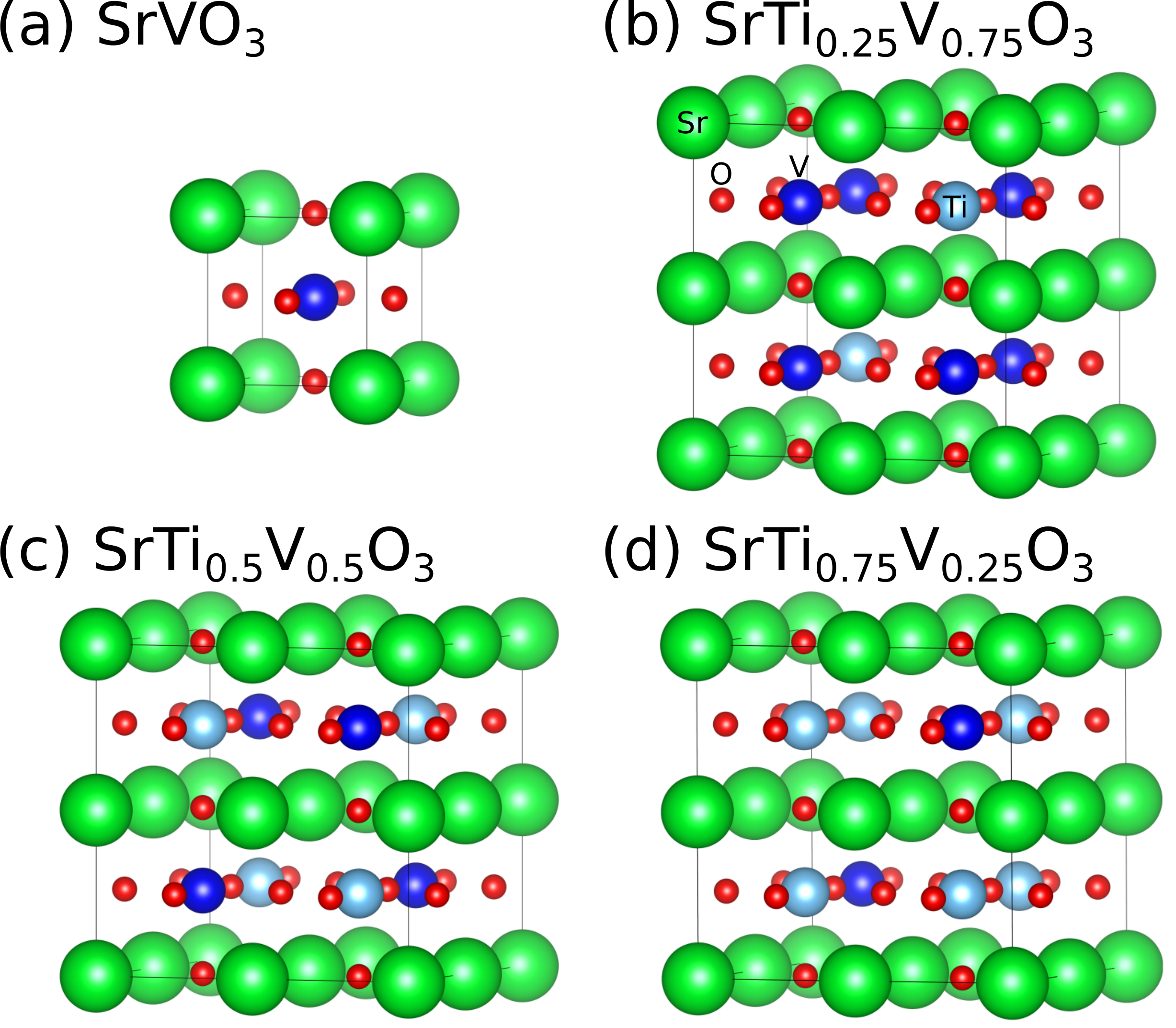}} 
 \caption{The unit cells of SrTi$_{\rm 1-x}$V$_{\rm x}$O$_3$ used in the supercell DFT and DFT+DMFT calculations. The configurations and stoichometries of SrTi$_{\rm 1-x}$V$_{\rm x}$O$_3$ are shown in (a) -- (d). The labels for each species are shown in (b). Bulk SrTiO$_3$ has the same crystal structure as SrVO$_3$ shown in (a).} 
\label{struct}
\end{figure}

DFT calculations were performed with the full potential augmented plane-wave plus local orbitals (APW+lo) {\sc elk} code~\cite{elk}, using SrTi$_{\rm 1-x}$V$_{\rm x}$O$_3$ cubic lattices shown in Fig.~\ref{struct} as input. These configurations were chosen to ensure inversion symmetry and  maximise V-V and Ti-Ti neighbour distances within the $2 \times 2 \times 2$ cell (other structures were also considered and are discussed in Appendix~\ref{appendix:a} and during the text). The DFT calculation used the Perdew-Burke-Ernzerhof (PBE) generalized gradient approximation (GGA) for the exchange-correlation functional~\cite{PhysRevLett.77.3865} and was converged on a $12 \times 12 \times 12 $ Monkhorst-Pack $\mathbf{k}$-mesh of 84 irreducible $\mathbf{k}$-points in the irreducible Brillouin zone. The (simple cubic) lattice parameter $a$ at which the DFT total energy is minimised for each of these structures was used. For comparison, additional virtual crystal approximation (VCA) calculations were performed at the same doping levels, ${\rm x}$, by replacing the V ion in a single SrVO$_3$ unit cell with an effective Ti$_{\rm 1-x}$V$_{\rm x}$ ion. We emphasise that VCA works well to describe compositional alloying and disorder of atoms which do not contribute to the states at the Fermi energy, and is not expected to perform well here.

\begin{figure}[t!]
 \centerline{\includegraphics[width=0.95\linewidth]{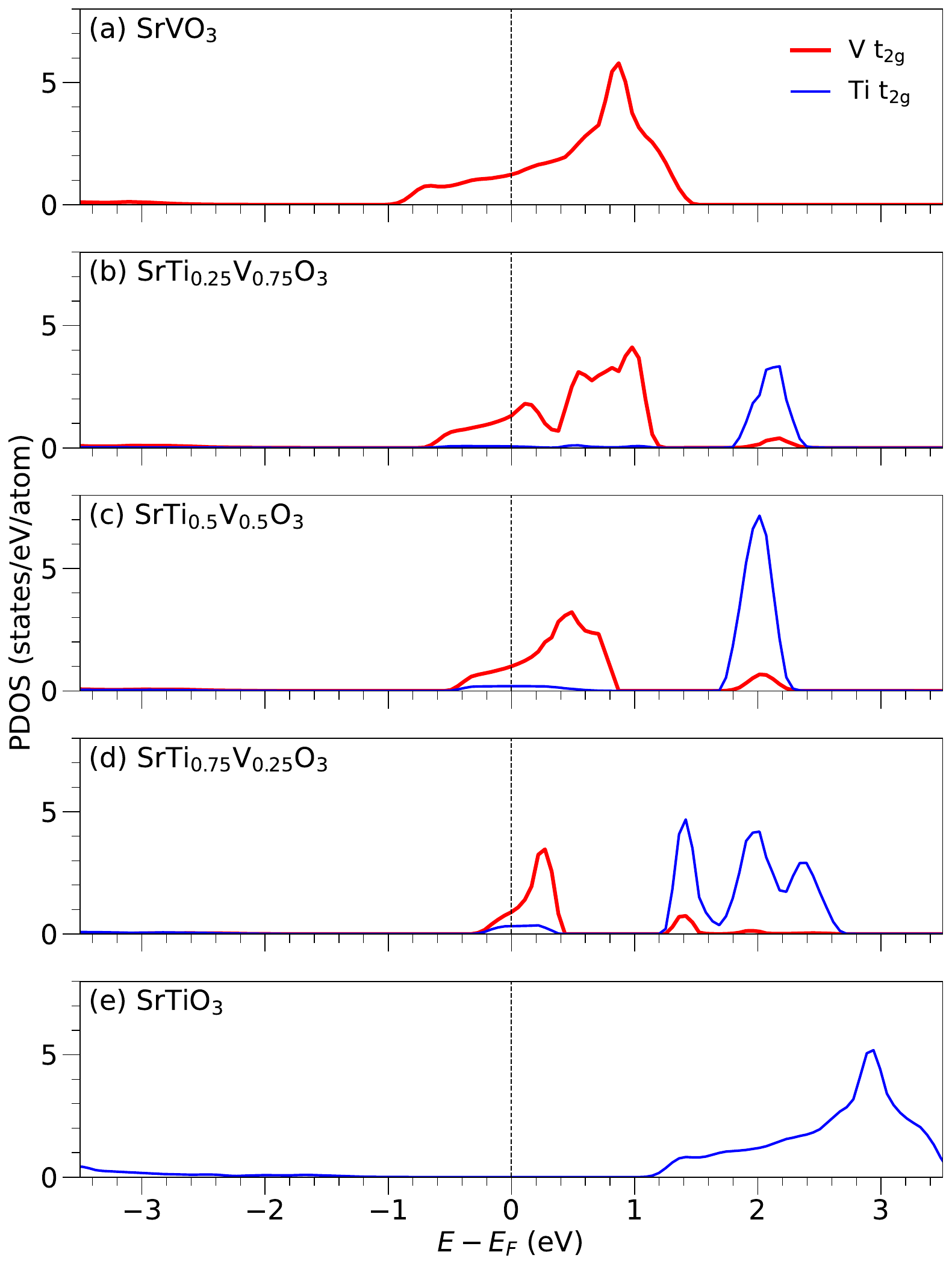}} 
 \caption{The DFT t$_{\rm 2g}$ partial density of states (PDOS) within the transition metal muffin-tin for each of the supercells shown in Fig.~\ref{struct}, including bulk (a) SrVO$_3$ and (e) SrTiO$_3$. At the DFT level, the Fermi level is located within the PDOS of the V t$_{\rm 2g}$ orbitals for the systems contains V, whereas it is within the band-gap for SrTiO$_3$.} 
\label{bulk-dft}
\end{figure}

For the DFT+DMFT calculations, the {\sc elk} code~\cite{elk} was used in combination with the toolbox for research on interacting quantum systems (TRIQS) library~\cite{triqs}. This so-called {\sc elk}+TRIQS DFT+DMFT framework is described in Ref.~\cite{james_2021}.
The DFT outputs were interfaced to the TRIQS/DFTTools application of the TRIQS library~\cite{aichhorn2016} by constructing Wannier projectors, as described in Ref.~\cite{james_2021},
derived from the bands around the Fermi level which are predominantly V t$_{\rm 2g}$ in character (see Fig.~\ref{bulk-dft}). The V t$_{\rm 2g}$ Wannier charge of each structure is 1 electron per V site. The DMFT part of the DFT+DMFT calculation was implemented using the continuous-time quantum Monte Carlo (CT-QMC) solver within the TRIQS/CTHYB application~\cite{seth2016} with the Kanamori-Hubbard interaction Hamiltonian parameterised by the Hubbard interaction $U = 4.0$~eV and Hund exchange interaction $J = 0.65$~eV, unless otherwise specified. We approximated the double counting within the fully localised limit. These $U$, $J$ values and double counting approximation are similar to those used for previous bulk SrVO$_3$ DFT+DMFT calculations~\cite{aichhorn2009}.  We used the fully-charge-self-consistent (FCSC) DFT+DMFT method with a total of $1.12\times10^{8}$ Monte Carlo sweeps within the impurity solver for each DMFT cycle. Owing to symmetry, there is only one impurity to solve for each doped system shown in Fig.~\ref{struct}. An inverse temperature $\beta = 40$ eV$^{-1}$ ($\sim$~290~K) was used. The spectral functions were calculated by analytically continuing the DMFT self-energy obtained from the {\em LineFitAnalyzer} technique of the maximum entropy analytic continuation method implemented within the TRIQS/Maxent application \cite{PhysRevB.96.155128}. 

In order to characterise the metal-insulator transition, certain quantities were extracted from the self-energies $\Sigma$ and Green's functions, which we define here for clarity. 
The presented spectral functions at the Fermi level are determined directly from the imaginary time Green's function by 
\begin{equation}
A(\omega=0) = \frac{\beta G(\tau=\frac{1}{2}\beta)}{\pi}, \label{e:A0}
\end{equation}
where $\beta$ is the inverse temperature in natural units. The $A(\omega=0)$ here is an averaged quantity over a frequency window
approximately equal to $\beta^{-1}$. These are normalised to the bulk SrVO$_3$ spectral function at the Fermi level to give $\overline{\rm A}$($\omega=0$). 
Our calculations (here and in Appendix~\ref{appendix:c}) where we refer to applying an effective potential (gate voltage V$_{\rm g}$) were achieved by providing a fractional excess electron charge to the $\rm{x}=0.5$ (and 0.25) supercell, resulting in a chemical potential shift of the uncorrelated states of eV$_{\rm g}$.

\section{Results}

\begin{figure}[t!]
 \centerline{\includegraphics[width=0.91\linewidth]{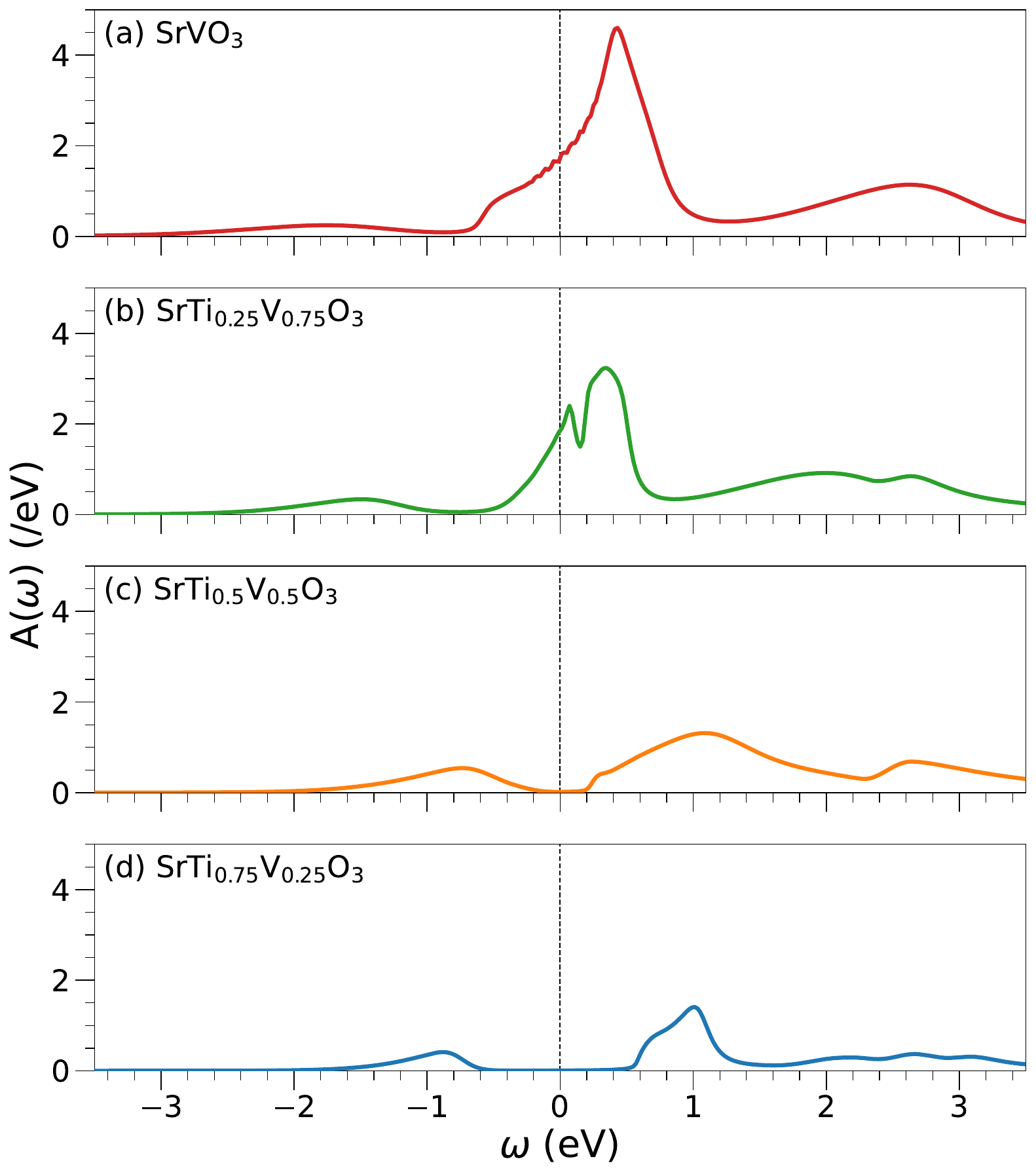}} 
 \caption{The DFT+DMFT integrated V Wannier spectral function A($\omega$) (including the V-Ti hybridised states in the Wannier projectors) of the supercells with decreasing V concentration from (a) to (d). The metal-insulator transition (MIT) occurs close to ${\rm x} \sim 0.5$. The Fermi level has been placed at the centre of the band-gap for the insulating systems.} 
\label{Aw-dmft-x}
\end{figure}

\begin{figure*}[t!]
 \centerline{\includegraphics[width=0.9\linewidth]{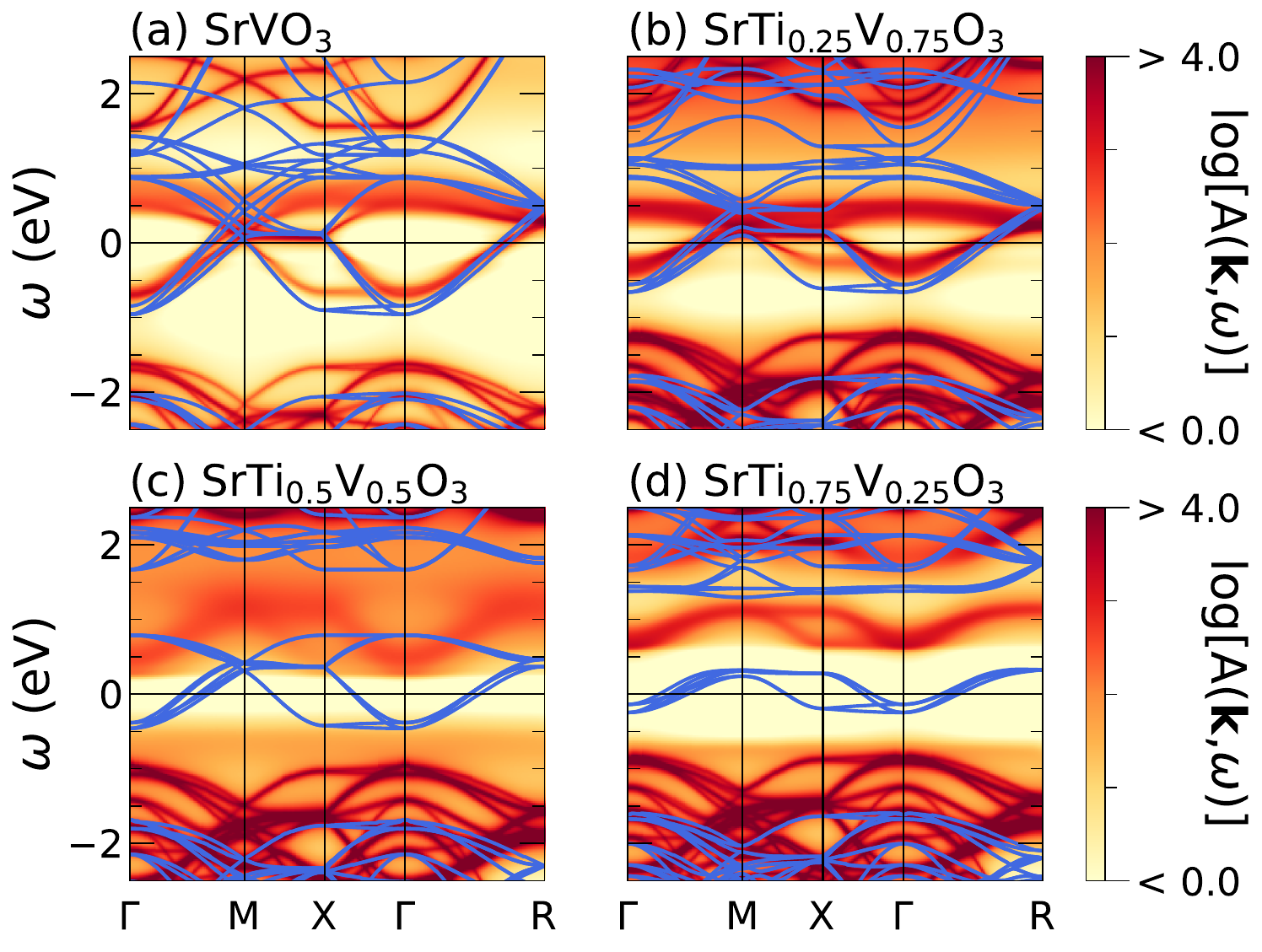}} 
 \caption{The DFT+DMFT $\textbf{k}$-resolved spectral function A($\textbf{k}$, $\omega$) of SrTi$_{\rm 1-x}$V$_{\rm x}$O$_3$, with decreasing V concentration from (a) to (d). Solid lines indicate the non-interacting DFT band structure. The bulk A($\textbf{k}$, $\omega$) and band structure in (a) have been back-folded into the $2 \times 2 \times 2$ supercell Brillouin zone.} 
\label{Akw-dmft-x}
\end{figure*}

The non-interacting t$_{\rm 2g}$ partial densities of states (PDOS) from DFT of SrTi$_{\rm 1-x}$V$_{\rm x}$O$_3$ are shown in Fig.~\ref{bulk-dft}. Except for SrTiO$_3$, which has no occupied electrons in the $t_{\rm 2g}$ manifold, all V-doped systems are metallic within DFT. The V t$_{\rm 2g}$ PDOS of the supercells in Figs.~\ref{bulk-dft} (b)--(d) clearly lie in an energy range near $E_{\rm F}$, which we refer to as the non-interacting bandwidth. The non-interacting V $t_{\rm 2g}$ bandwidth decreases as the V concentration $\rm x$ is decreased, dropping by a factor of 3 between x = 1 and x = 0.25. There are additional contributions in the Ti and V t$_{\rm 2g}$ PDOS which arise due to hybridisation between these orbitals. We note that in the DFT calculations the V t$_{\rm 2g}$ PDOS is the largest contribution to the states at and around the Fermi level in all the systems and configurations investigated. Although DFT adequately qualitatively describes the quasiparticle peak of the V t$_{\rm 2g}$ states in SrVO$_3$, it fails to capture the quasiparticle bandwidth and the Hubbard bands due to its neglect of electron-electron interactions, and no MIT is observed at the level of DFT for the doped systems. We note that recent DFT+U calculations of SrTi$_{\rm 0.75}$V$_{\rm 0.25}$O$_3$ in the low-temperature (orthorhombic) structure of SrTiO$_3$ were shown to be insulating due to distortion-induced crystal-field effects~\cite{georgescu2023cudoped}.

Figure \ref{Aw-dmft-x} shows the results of supercell DFT+DMFT calculations, where the Wannier V DOS is shown for each doped system. In agreement with previous calculations \cite{nekrasov2005,aichhorn2009} the DFT+DMFT correctly predicts the Hubbard bands for pure SrVO$_3$ (Fig.~\ref{Akw-dmft-x}), which appear at $\approx -1.7$~eV (lower Hubbard band, LHB) and $\approx 2.7$~eV (upper Hubbard band, UHB). As the Ti concentration is increased, the V Hubbard band intensity rises as spectral weight is transferred from the quasiparticle peak, and the bandwidths of the V quasiparticle states are renormalised with respect to the non-interacting (DFT) bandwidth (see Fig.~\ref{Akw-dmft-x}). At a critical doping, $\rm{x}_{\rm c} \sim 0.5$, the system undergoes a metal-insulator transition as the spectral weight is completely diminished at the Fermi level. This is in very good agreement with previous experiments. 
Similar results are obtained with a different value of $U$ and for different atomic arrangements of the supercell (see Appendix~\ref{appendix:a} and following discussion). We note that the atomic configuration of the $\rm{x}=0.75$ supercell employed here lifts the degeneracy of the V t$_{\rm 2g}$ orbitals, which gives rise to a small dip in the V DOS in Fig.~\ref{Aw-dmft-x}(b); this is an artefact of the supercell approach. 
Amongst the upper Hubbard bands of the doped systems, there are additional distinct broad features above about 2.5 eV. These correspond to hybridised V t$_{\rm 2g}$ and Ti t$_{\rm 2g}$ states as seen in the DFT PDOS in Fig.~\ref{bulk-dft}.

\begin{figure}[t!]
 \centerline{\includegraphics[width=0.995\linewidth]{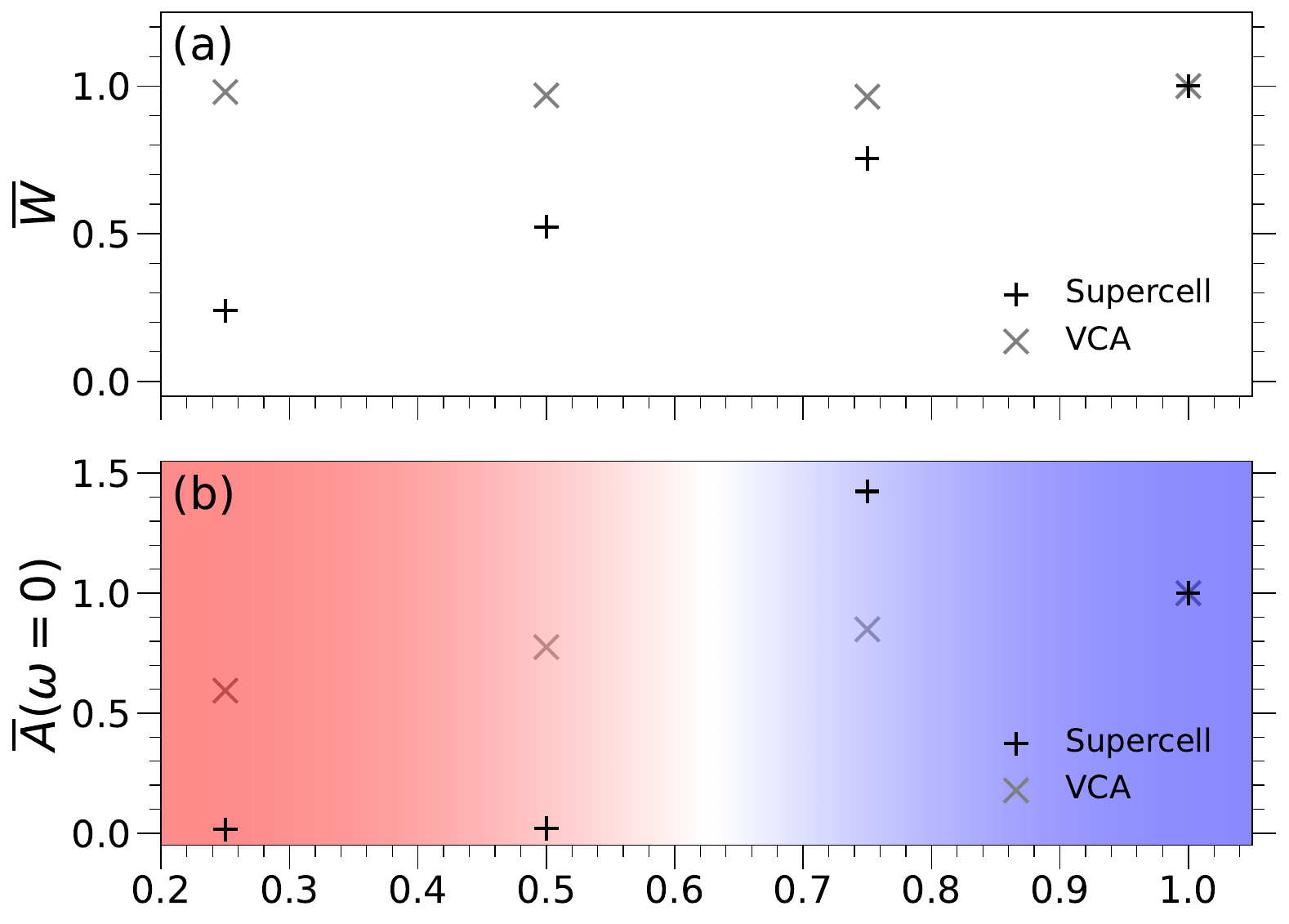}} 
 \caption{(a) The VCA and supercell DFT (normalised) orbital-averaged bandwidths $\overline{W}$ of the V t$_{\rm 2g}$ bands located around the Fermi level (which relate to the bandwidths of the V t$_{\rm 2g}$ PDOS shown in Fig.~\ref{bulk-dft}). (b) The VCA and supercell DFT+DMFT (normalised) integrated V Wannier spectral function at the Fermi level $\overline{\rm A}$($\omega=0$) showing the metal-insulator transition near $\rm{x}=0.5$. The overline over the variables shown in this figure refers to the quantity being normalised to the corresponding bulk SrVO$_3$ value. These quantities highlight the reduced coherence of the V t$_{\rm 2g}$ states near the transition close to $\rm{x}=0.5$. The red (left) to blue (right) background colour of (b) is a guide to the eye of which regions are insulating (left) and metallic (right).}
\label{dmft-x-comp}
\end{figure}

As illustrated in Figs.~\ref{Akw-dmft-x} and \ref{dmft-x-comp}~(a), all doped structures are metallic from the DFT predictions, which clearly shows that (near $\rm{x}_{\rm c}$) the insulating phase is Mott in nature, arising from the local electron correlations included in DMFT. The supercell non-interacting (orbital-averaged) bandwidth from bare DFT ($\overline{W}$) is shown in Fig.~\ref{dmft-x-comp}~(a), and decreases almost linearly from SrVO$_3$ to SrTi$_{0.75}$V$_{0.25}$O$_3$. This reduction in the quasiparticle bandwidth is greater than that achieved in other doped SrVO$_3$ systems, such as Sr$_{\rm x}$Ca$_{\rm 1-x}$VO$_3$~\cite{laverock2015b,laverock2013b}. Correspondingly, the integrated spectral function at the Fermi level $\overline{\rm A}$($\omega=0$) [displayed in Fig.~\ref{dmft-x-comp}~(b)] increases as the quasiparticle peak width shrinks as the doping approaches $\rm{x}_{\rm c}$, indicating the increasing influence of the local electron correlations within the t$_{\rm 2g}$ orbitals (and has been seen for tuning $U$ through the MIT in SrVO$_3$ superlattice heterostructures~\cite{james2021quantum}). These can be explained by the $U$/$W$ ratio increasing when approaching the MIT from high doping in the metallic phase. In the insulating phase, $\overline{\rm A}$($\omega=0$) is equal to zero as expected. 
Figs.~\ref{dmft-x-comp}~(a) and (b) show our VCA DFT and DFT+DMFT calculations used as an alternative to approximate the effect of the substitutional doping. However, Fig.~\ref{dmft-x-comp}~(b) shows that these calculations do not give rise to an MIT. Here, the effect of the VCA is to effectively shift the t$_{\rm 2g}$ states (with respect to the Fermi level) to conserve charge, and hence the almost constant $\overline{W}$ with respect to doping in Fig.~\ref{dmft-x-comp}~(a). The VCA unsurprisingly does not capture the variations in the potentials around each transition metal site which evidently needs to be included.

As with the results of monolayer~\cite{bhandary2016}, bilayer~\cite{zhong2015} or superlattice heterostructures~\cite{james2021quantum} of SrVO$_3$, the spectral weight from the V t$_{\rm 2g}$ states in the $\rm{x}=0.25$ and $0.5$ doped systems appears to fully redistribute to the Hubbard bands in this insulating phase. The UHB appears to have a momentum dependent dispersion in Fig.~\ref{Akw-dmft-x} (c) and (d), similar to that seen in Ref.~\cite{james2021quantum}. This dispersive UHB, which is strongly reminiscent of the underlying QP dispersion, is very prominent in the A($\textbf{k}$, $\omega$) of $\rm{x}=0.25$ system. DFT+DMFT calculations using a real-time impurity solver have shown that the UHB of bulk SrVO$_3$ is composed of three atomic states (namely the excited two electron [$N=2$] atomic states)~\cite{PhysRevX.7.031013}. The ill-posed nature of analytical continuation results in the finer details of these atomic states being ``washed out'' to give the typical broad UHB feature seen in bulk SrVO$_3$~\cite{PhysRevX.7.031013} (see Fig.~\ref{Aw-dmft-x} (a)). In our results, these atomic states appear to be more prominent in the spectral function derived from the analytically-continued self-energy for the low $\rm{x}$ doped systems (also see Appendix~\ref{appendix:b}). Indeed, the shouldered feature around 1 eV in the $\rm{x}=0.25$ system appears to be composed of the two singly occupied $N=2$ atomic states (see Appendix~\ref{appendix:b}). 


\begin{figure}[t!]
 \centerline{\includegraphics[width=0.995\linewidth]{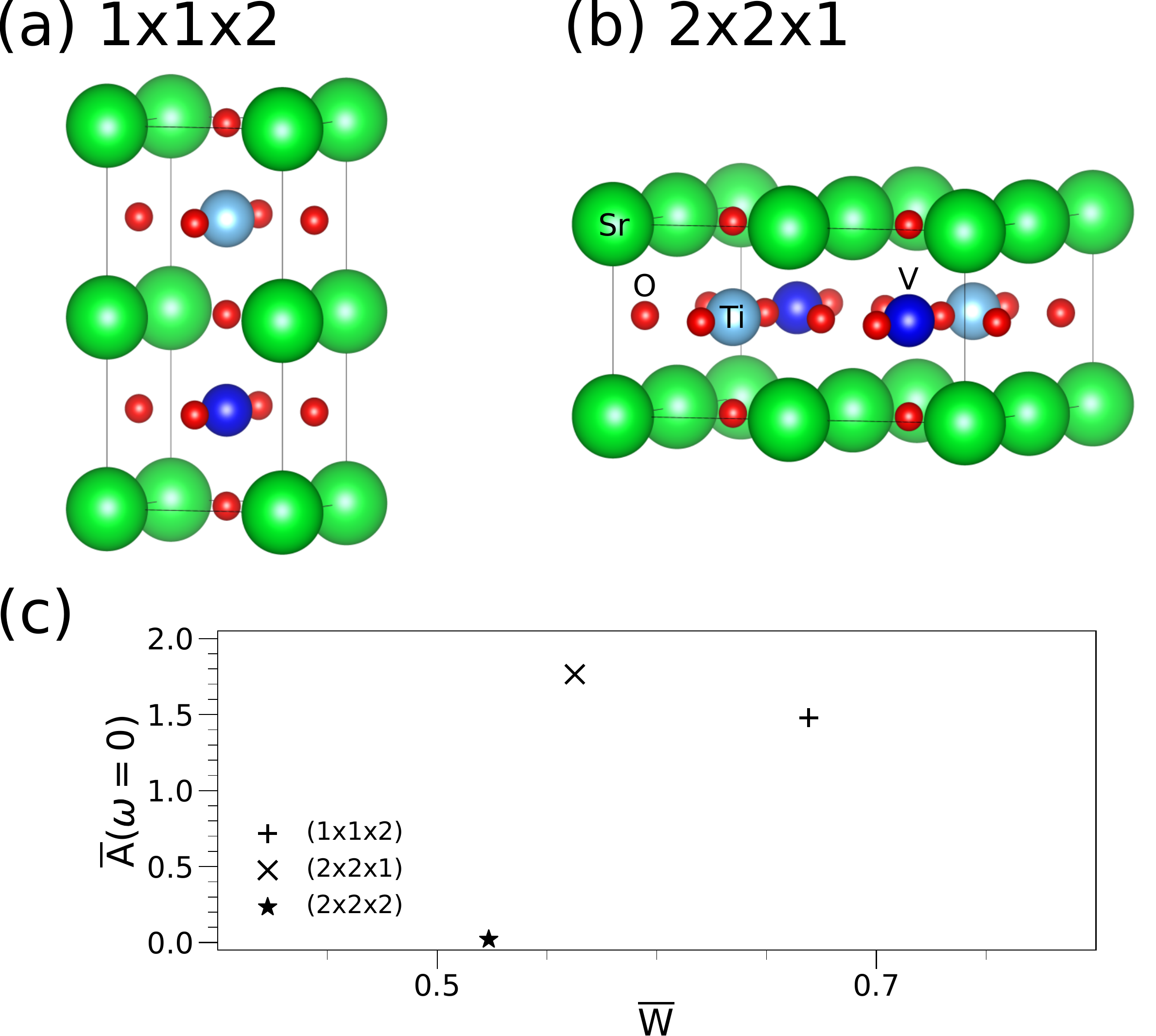}} 
 \caption{The (a) 1$\times$1$\times$2 and (b) 2$\times$2$\times$1 supercell configurations investigated for $\rm{x}=0.5$ doping. The labels for each species are shown in (b). (c) $\overline{\rm A}$($\omega=0$) 
 (as defined in Fig.~\ref{dmft-x-comp}) for different V and Ti site configurations within different unit cells of SrTi$_{\rm 0.5}$V$_{\rm 0.5}$O$_3$, given as a function of the corresponding normalised DFT bandwidth $\overline{W}$. The red (left) to blue (right) background colour of (c) is a guide to the eye of which regions are insulating (left) and metallic (right).} 
\label{dmft-x0.5-comp}
\end{figure}

The impact of atomic configuration is clearly shown for $\rm{x}=0.5$ in Fig.~\ref{dmft-x0.5-comp}. In this figure, three different unit cells of $\rm{x}=0.5$ are presented with the V atoms having the shortest average intra-species distance in the 1$\times$1$\times$2 unit cell (Fig.~\ref{dmft-x0.5-comp}~(a)), the longest average intra-species distance in the 2$\times$2$\times$2 (i.e., the structure in Fig.~\ref{struct}~(c)) and the average intra-species distance in the 2$\times$2$\times$1 unit cell [Fig.~\ref{dmft-x0.5-comp}~(b)] is inbetween the other two. The corresponding results of these unit cells presented in Fig.~\ref{dmft-x0.5-comp}~(c) show that the $\rm{x}=0.5$ is at the precipice of the MIT and is highly dependent on the position configurations of the transition metal species. It also clearly indicates the critical normalised non-interacting orbital-averaged bandwidth value $\overline{W}_{\rm c}$  (which is around 0.55~eV) at which the MIT occurs. 
All of the $\rm{x}=0.5$ configurations can be driven into the insulating phase with a modest increase in $U$ to 5~eV (while $\rm{x} = 0.75$ remains metallic). Experimentally, the rather broad range of $\rm{x}_{\rm c}$ that is reported (between 0.4 and 0.7) is also influenced by other factors beyond disorder such as  the different substrates (e.g.\ SrTiO$_3$~\cite{Kanda_2021} or LSAT~\cite{Gu_2013_STVO}) used, film thicknesses, and measuring bulk crystals~\cite{Tsuiki_1983,HONG2002305}. In particular, SrTiO$_3$ and LSAT substrates introduce quite different in-plane strain to the SrTi$_{\rm 1-x}$V$_{\rm x}$O$_3$ lattice in epitaxial thin films, so that crystal-field effects become important near $\rm{x}_{\rm c}$.

\begin{figure}[t!]
 \centerline{\includegraphics[width=0.995\linewidth]{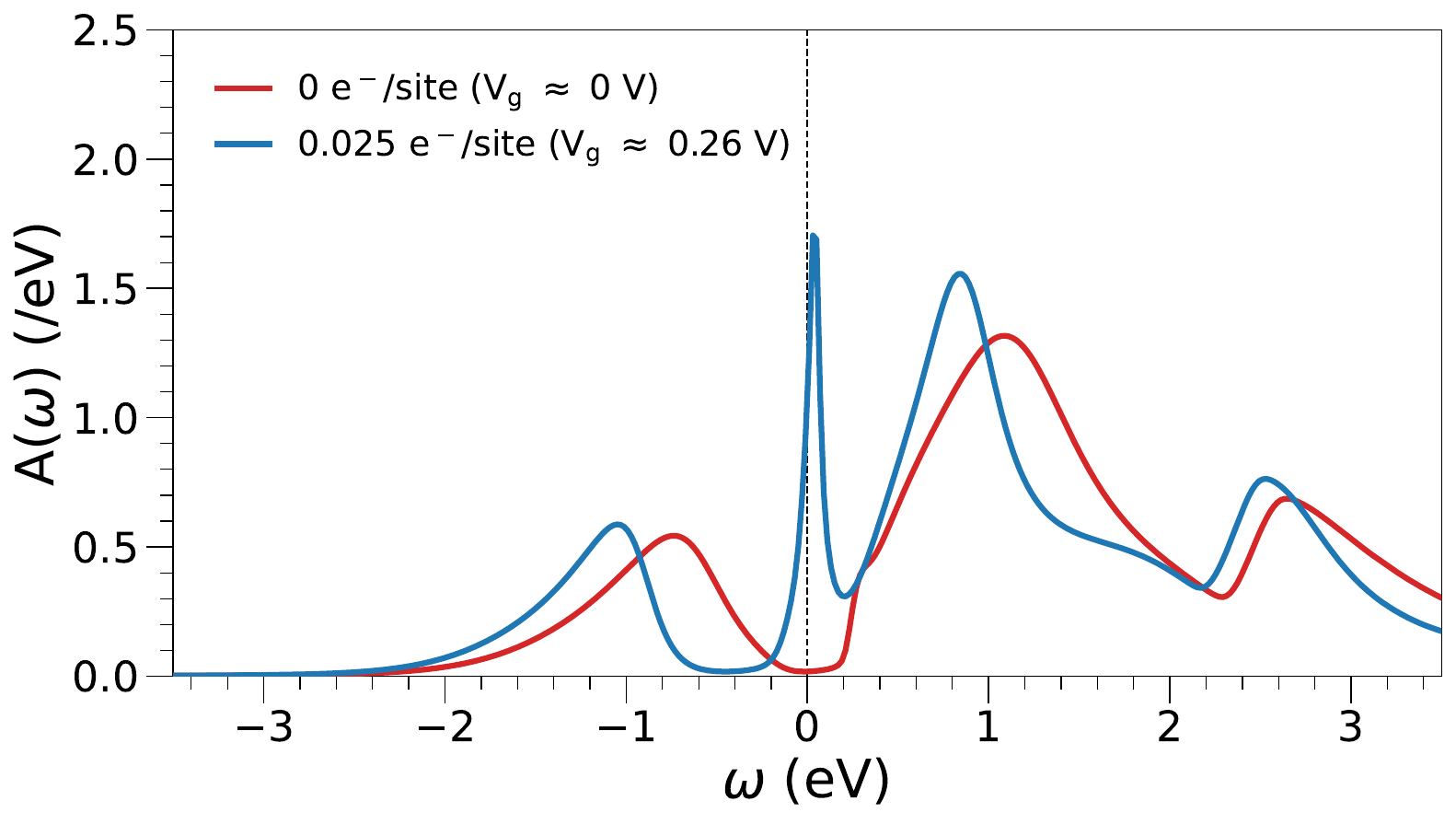}} 
 \caption{The Wannier V t$_{\rm 2g}$ A($\omega$) with the V-Ti hybridised states of the $\rm{x}=0.5$ system [in Fig.~\ref{Aw-dmft-x} (c)] and of the same system but doped with an additional 0.025 of an electron per V site (e$^-$/site). This is equivalent to a gating voltage V$_{\rm g}$ of 0.26~V. This doping pushes the system into being metallic showing that external perturbations with the same effect can switch the $\rm{x}=0.5$ system between the metallic and insulator state. This gives the possibility of SrTi$_{\rm 1-x}$V$_{\rm x}$O$_3$ being used for Mott electronic devices. The Fermi level has been placed at the centre of the band-gap for the insulating system.} 
\label{x0.5-chgexs}
\end{figure}

Finally, we show that these substitutionally-doped SrTi$_{\rm 1-x}$V$_{\rm x}$O$_3$ system near $\rm{x}_{\rm c}$ can be switched between the metallic and insulating state in Fig.~\ref{x0.5-chgexs}. By applying an external (gating) potential V$_{\rm g}$ (which gives the equivalent effect as the charge excess in these charge excess calculations) in the $\rm{x}=0.5$ system, this results in it transitioning from the Mott insulating phase to the metallic one. The V$_{\rm g}$ for $\rm{x}=0.5$ is similar to that in the bilayer system presented by Zhong $\textit{et al.}$~\cite{zhong2015}. A larger potential is needed for the $\rm{x}=0.25$ system (discussed in Appendix~\ref{appendix:c}) as expected as it is further away from $\rm{x}_{\rm c}$. This shows that these systems could be used in Mott devices with this switching mechanism controlled by an appropriate external perturbation. As these substitutionally-doped systems are bulk in nature, these will not be prone to issues associated with maintaining a pristine surface as would be required for the bilayer-capped SrVO$_3$ system presented by Zhong $\textit{et al.}$~\cite{zhong2015}. 

Given our DFT+DMFT supercell results, it is clear that more detailed investigations near $\rm{x}_{\rm c}$ will be invaluable. Computationally, calculations of doping levels at a finer granularity than presented here remain out of reach using an all-electron supercell approach. However, new methods beyond DFT+DMFT that are appropriate for disordered systems are rapidly becoming available. For example, Refs.~\cite{Nguyen_2022,Weh_2021} describe new approaches that are able to capture the local atomic configuration effects as well as incorporating the additional Anderson localisation effects due to site-disorder for interacting electron systems. Features in the spectral function such as localised Anderson states (originating from disorder partitioning the electron states into either localised or extended states) within the Mott insulating gap are present in such models (e.g., see Ref.~\cite{Nguyen_2022}) which are not captured in the supercell results presented here. 
With the clear dependence on the site configurations and the local electron correlation effects of the V t$_{\rm 2g}$ states around the Fermi level, SrTi$_{\rm 1-x}$V$_{\rm x}$O$_3$ could serve as a prototypical test material for initial real-material calculations of solving such Anderson-Mott systems, just as SrVO$_3$ was an initial focus when developing DFT+DMFT.

\section{conclusion}
We have shown that the measured trends around the MIT in the doped SrTi$_{\rm 1-x}$V$_{\rm x}$O$_3$ system are captured well by DFT+DMFT predictions of different supercell configurations. Refinements in both DFT and DMFT implementations allows these supercell DFT+DMFT calculations to be more computational feasible since when some of these ARPES measurements were published about a decade ago. From our supercell results, it is clear that both the local electron correlations and site configurations are vital for this MIT. It is the site configurations which influences the non-interacting bandwidth, which in combination with the electron correlation effects from the onsite Coulomb interaction gives rise to the MIT. This therefore is a likely candidate for a Mott-Anderson MIT. We also note that atomic states associated to the UHB are more prominent in these doped systems, whereas in bulk SrVO$_3$ these atomic states are ``washed out'' as a consequence of analytic continuation, yielding the familiar broad UHB feature.

By providing an external perturbation, equivalent to the effect of an additional small excess of electron doping, SrTi$_{\rm 1-x}$V$_{\rm x}$O$_3$ can be made to switch between the Mott insulating and metallic phases, giving another possible avenue to develop Mott electronic devices. These results scratch the surface of the complex strong local correlated electron behaviour meaning that further work is needed to probe this possible Mott-Anderson MIT to understand how the different factors such as site disorder and the local electron correlations influences the MIT around the critical doping which is needed to realise its potential for Mott devices.

\section{Acknowledgements}
A.D.N.J. acknowledges the Doctoral Prize Fellowship funding and support from the Engineering and Physical Sciences Research Council (EPSRC). 
Calculations were performed using the computational facilities of the Advanced Computing Research Centre, University of Bristol (\href{http://bris.ac.uk/acrc/}{http://bris.ac.uk/acrc/}). The VESTA package (\href{https://jp-minerals.org/vesta/en/}{https://jp-minerals.org/vesta/en/}) has been used in the preparation of some figures.

\appendix
\counterwithin{figure}{section}

\section{Unit cell configuration}
\label{appendix:a}

\begin{figure}[t]
 \centerline{\includegraphics[width=0.9\linewidth]{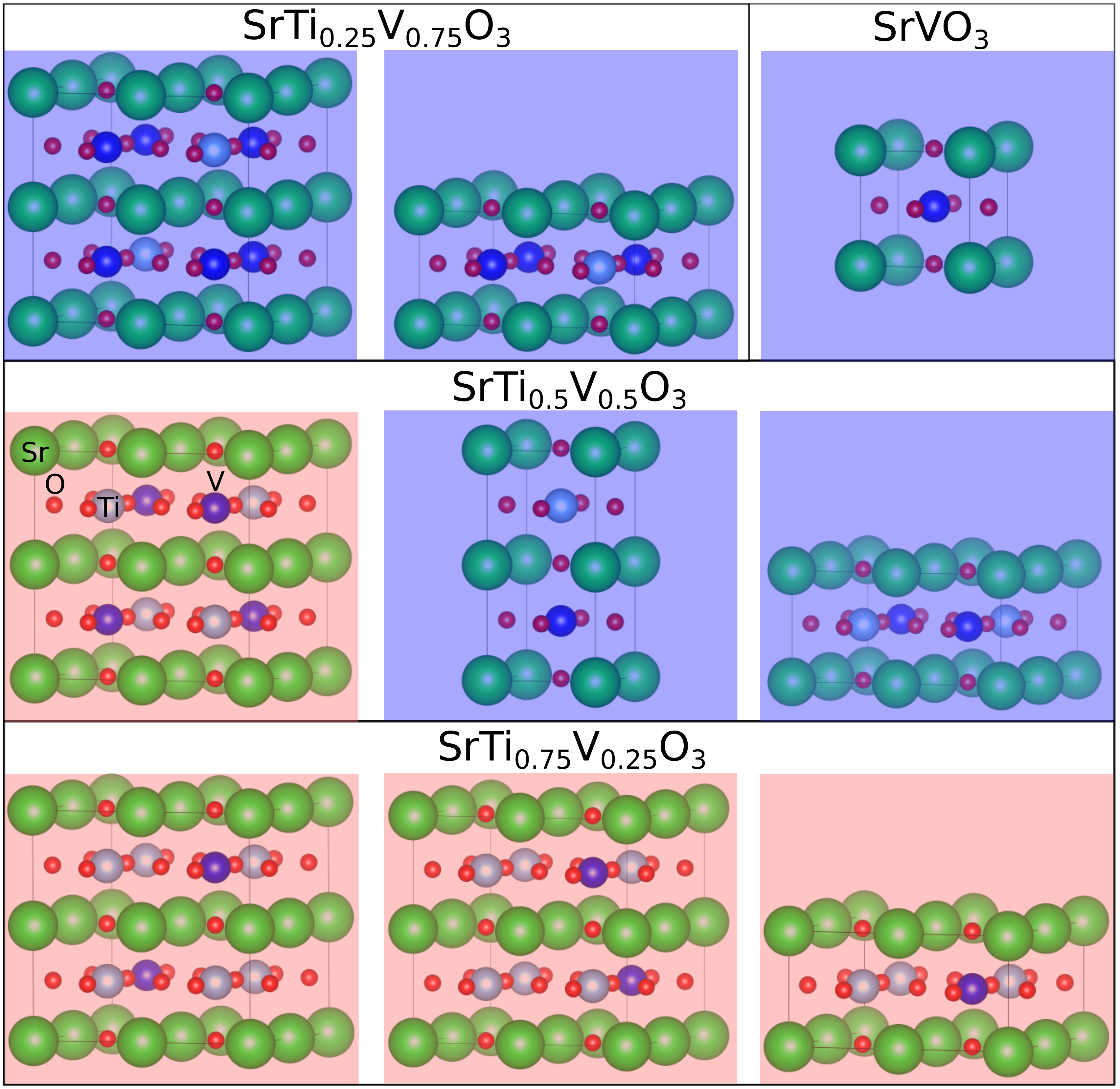}} 
 \caption{The different investigated SrTi$_{\rm 1-x}$V$_{\rm x}$O$_3$ configurations up to 2x2x2 sized unit cells for $\rm{x}=0.25$, 0.5 and 0.75. The SrVO$_3$ unit cell is included for completeness. Each configuration is masked with an translucent box which indicates whether the DFT+DMFT $U = 4.0$~eV results of that configuration is metallic (light blue) or insulating (light red), for example, the SrVO$_3$ unit cell is metallic.}
\label{S1}
\end{figure}

Figure~\ref{S1} shows all the unit cell configurations (up to 2$\times$2$\times$2) used for this work. The lattice parameter was determined by minimising the total DFT energy as discussed in the main text. For the $U = 4.0$~eV DFT+DMFT calculations, each configuration for the $\rm{x}=0.25$ doped systems are all insulating whereas all of the $\rm{x}=0.75$ (and 1.0) configurations are metallic. For the $\rm{x}=0.5$ configurations, we see that the metallic/insulating phase is dependent on the configuration (which influences the DFT ``non-interacting'' bandwidths $W$ and as such on the electron correlations which are usually quantified by the $U$/$W$ ratio $U$/$W$ ratio) as discussed in the main text. We find that using $U = 5.0$~eV pushes all of the $\rm{x}=0.5$ configurations insulating as a consequence of its effect on the electron correlations seen by the $U$/$W$ ratio. We note that the band gap in the insulating phase increases as the $W$ decreases, again as a result of enhancement of local correlation quantified by the $U$/$W$ ratio.

\section{$U$-varied Metal-Insulator Transition}
\label{appendix:b}

\begin{figure}[t]
 \centerline{\includegraphics[width=0.7\linewidth]{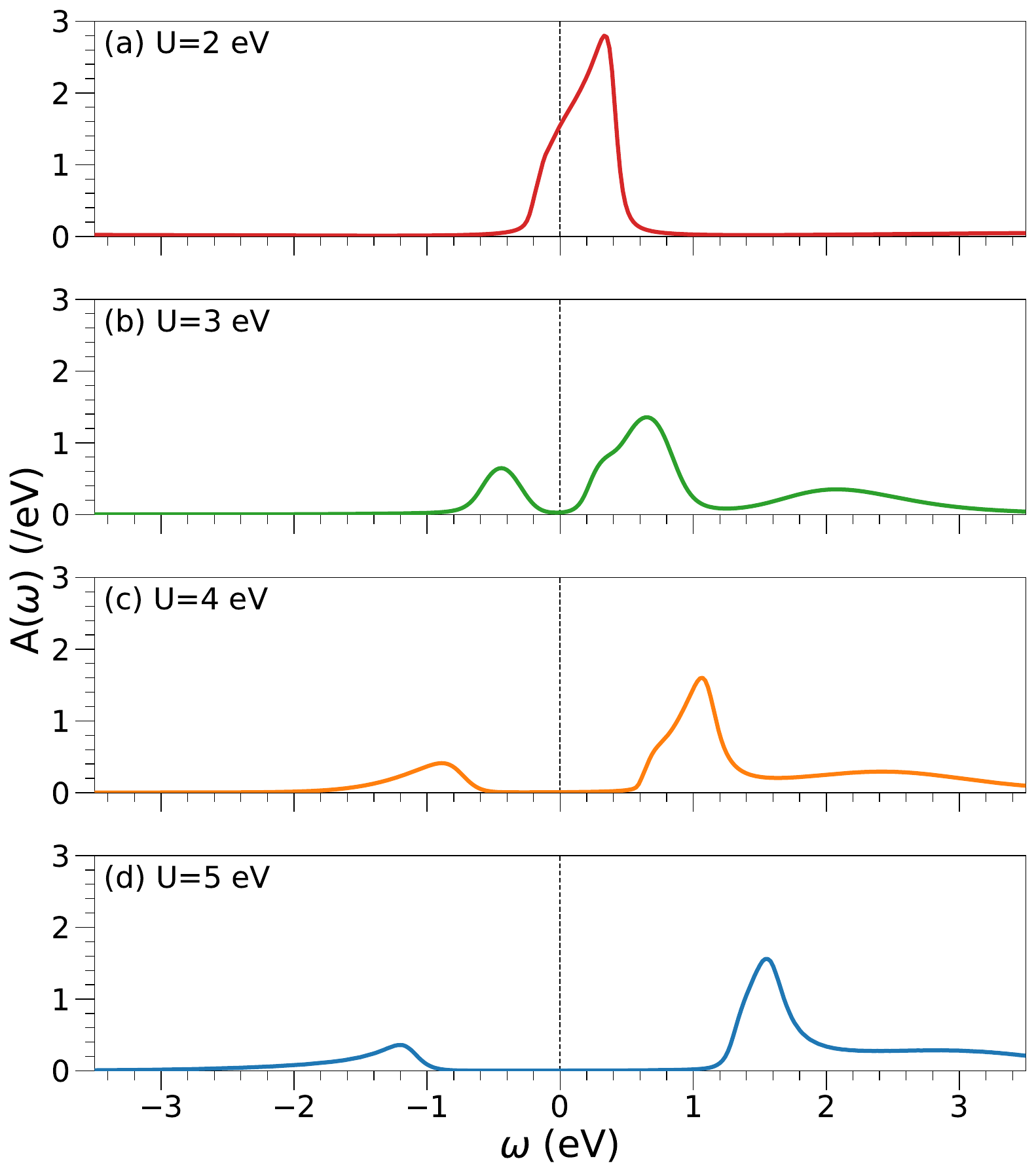}} 
 \caption{The effect of varying $U$ (for fixed $J=0.65$ eV) on the $\rm{x}=0.25$ Wannier V t$_{\rm 2g}$ spectral functions (calculated from the Wannier projectors without the Ti-V hybridised states) through the metal-insulator transition. The Fermi level has been placed in the centre of the band-gap for the insulating results.}
\label{S2}
\end{figure}

\begin{figure}[t]
 \centerline{\includegraphics[width=0.7\linewidth]{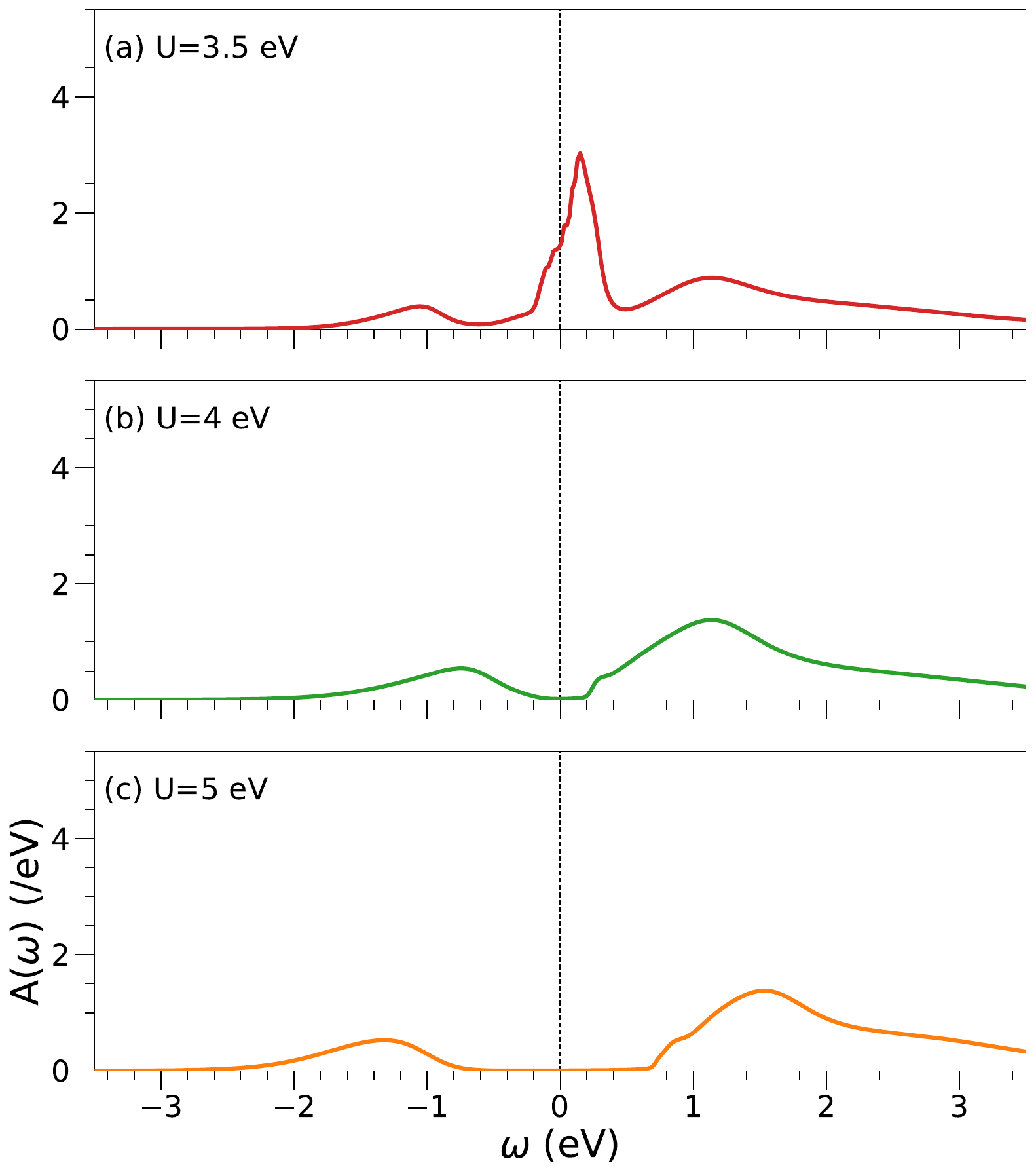}} 
 \caption{The effect of varying $U$ (for fixed $J=0.65$ eV) on the $\rm{x}=0.5$ Wannier V t$_{\rm 2g}$ spectral functions (calculated from the Wannier projectors without the Ti-V hybridised states) through the metal-insulator transition. The Fermi level has been placed in the centre of the band-gap for the insulating results.}
\label{S3}
\end{figure}

We also show the metal-insulator transition (MIT) from varying the on-site Hubbard $U$ (for fixed $J=0.65$ eV) in Figs.~\ref{S2} and \ref{S3} for doping $\rm{x}=0.25$ and $0.5$ (of the unit cell configuration in the main text), respectively. We see from the Wannier spectral functions in Figs.~\ref{S2} and \ref{S3} (which are calculated from using the Wannier projectors constructed from the V t$_{\rm 2g}$ bands at the Fermi level) that the spectral weight of the quasiparticle states is totally depleted and redistributed to the upper and lower Hubbard bands for both the $\rm{x}=0.25$ and $0.5$ doped systems. The Mott insulating band gap increases with $U$ as expected for these Hubbard bands. 

Interestingly, in the Mott insulating phase of the $\rm{x}=0.25$ system Fig.~\ref{S2} (b)-(d), there are two distinct features above the Fermi level, both of which we associate to the upper Hubbard band. Here, the higher energy feature is broad in nature (and is indistinguishable amongst the V-Ti hybridised states in the main text) whereas the lower energy feature has a lower energy shoulder which becomes suppressed with increasing $U$ until the shoulder is no longer visible for $U=5$ eV. The (peak-to-peak) energy separation between these two features of the upper Hubbard band remains constant as a function of $U$. These properties of these two observed features of the upper Hubbard band ties in well with the observed three $N=2$ (two electron) excited atomic states calculated for bulk SrVO$_3$ in Ref.~\cite{PhysRevX.7.031013}. The three $N=2$ atomic states are: same spin singly occupied orbitals (lowest energy relative to the ground state); different spin singly occupied orbitals; and doubly occupied orbitals (highest energy relative to the ground state)~\cite{PhysRevX.7.031013}. Therefore the shoulder is likely associated with the same spin singly occupied atomic state, the peak of the lower feature of the upper Hubbard band relates to the different spin singly occupied atomic state, which leaves the higher energy feature of the upper Hubbard band being connected to the doubly occupied atomic state. In Fig.~\ref{S2} (b) and (c) The energy difference between the two singly occupied orbital states is much smaller than between a singly occupied orbital state and the doubly occupied atomic state, which agrees with the atomic states seen in bulk SrVO$_3$ and the solution to the atomic problem of the interaction Hamiltonian presented in Ref.~\cite{PhysRevX.7.031013}. All of this indicates that these spectral function features above the Fermi level are likely associated to the $N=2$ atomic states. Analytic continuation is likely smearing out the finer details of these atomic states in the upper Hubbard band which is why the shoulder is being suppressed assuming the spectral weight around the $N=2$ same spin singly occupied atomic state changes for higher $U$ values as seen in Ref.~\cite{PhysRevX.7.031013}.

For the $\rm{x}=0.5$ system in Fig.~\ref{S2}, the upper Hubbard band is a more broad feature similar to bulk SrVO$_3$. The atomic states are less distinguishable but there is a notable smoothed kink in the spectral function above the peak upper Hubbard band energy (e.g., at about 1.8 eV in Fig.~\ref{S2} (b)) which may also indicate the $N=2$ atomic states, where the $N=2$ singly occupied atomic states are still likely at lower energies than the doubly occupied atomic state.

\section{External-Potential-driven Metal-Insulator Transition}
\label{appendix:c}

\begin{figure}[t]
 \centerline{\includegraphics[width=0.7\linewidth]{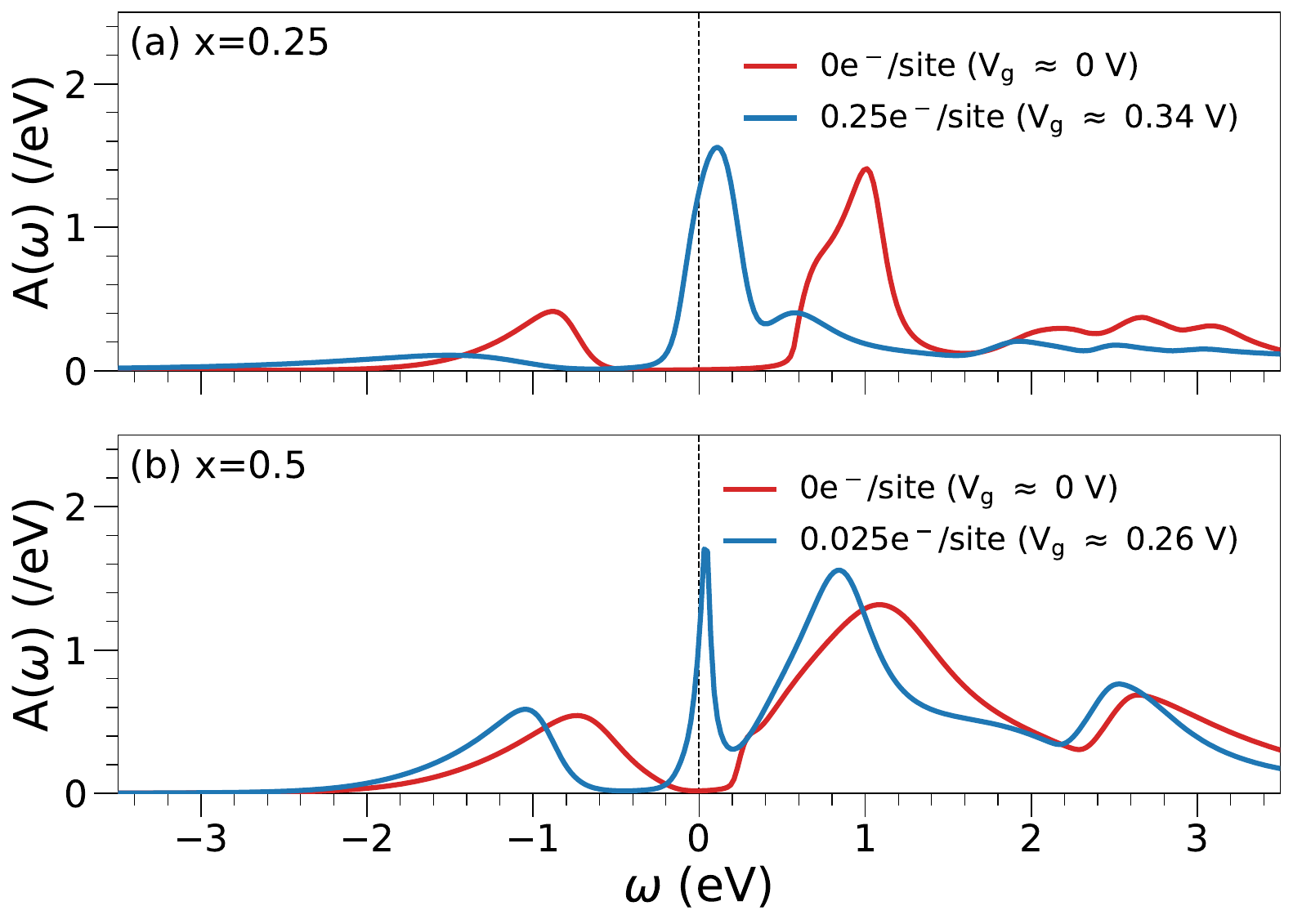}} 
 \caption{The effect of varying applying an effective external (gating) potential V$_{\rm g}$ on the $\rm{x}=0.25$ and 0.5 Wannier V t$_{\rm 2g}$ spectral functions (calculated from the Wannier projectors without the Ti-V hybridised states) which pushes the insulating state through the metal-insulator transition to a clear metallic state. The effect of V$_{\rm g}$ (0.34~V for $\rm{x}=0.25$ and 0.26~V for $\rm{x}=0.5$) is equivalent to an excess electron doping of 0.25 per V site (e$^-$/site) for $\rm{x}=0.25$ and 0.025 per V site (e$^-$/site) for $\rm{x}=0.5$. The Fermi level has been placed in the centre of the band-gap for the insulating results.}
\label{S4}
\end{figure}

In the main text, we discuss the influence of applying an external potential to the $\rm{x}=0.5$ system (of the unit cell configuration in the main text) which pushes it from the insulating phase to the metallic one. This is shown in Fig.~\ref{S4} along with the larger potential needed to push the $\rm{x}=0.25$ insulating system (of the unit cell configuration in the main text) to become metallic. These DFT+DMFT results are from the calculations using $U = 4.0$~eV. These results show that the MIT from an external potential is not necessarily confined to be around the critical doping $\rm{x}_{\rm c}$, possibly giving more flexibility for use in Mottronic devices. 

\bibliography{ref_new}

\end{document}